# PERFORMANCE IMPROVEMENT OF AN AS-FRIENDLY PEER SELECTION ALGORITHM FOR P2P LIVE STREAMING

Yukinobu Fukushima, Kazuki Kobayashi and Tokumi Yokohira

Graduate School of Natural Science and Technology,
Okayama University, Tsushima-naka 3-1-1, Kita-ku, Okayama-city, Okayama, Japan

***ABSTRACT***

*Minimum Physical Hop (MPH) has been proposed as a peer selection algorithm for decreasing inter-AS (Autonomous System) traffic volume in P2P live streaming. In MPH, a newly joining peer selects a peer whose physical hop count (i.e., the number of ASes traversed on the content delivery path) from it is the minimum as its providing peer. However, MPH shows high inter-AS traffic volume when the number of joining peers is large. In this paper, we propose IMPH that tries to further decrease the inter-AS traffic volume by distributing peers with one logical hop count (i.e., the number of peers or origin streaming servers (OSSes) traversed on the content delivery path from an OSS to the peer) to many ASes and encouraging the following peers to find their providing peers within the same AS. Numerical examples show that IMPH achieves at the maximum of 64% lower inter-AS traffic volume than MPH.*

## 1. INTRODUCTION

There has been a big demand for live streaming services over the Internet. As one of the service models, a peer-to-peer (P2P) live streaming system based on a P2P model has been attracting attention [1-5, 19, 20]. The system can mitigate the load of Origin Streaming Servers (OSSes), which provide an original content data, by making user terminals (peers) that have already joined the system provide the content data for other peers in a multihop manner.

In P2P live streaming systems, it is important to 1) reduce inter-AS (Autonomous System) traffic volume for restricting transit costs for ASes [6] and 2) keep the real-time property of content for realizing live streaming [7, 8]. Inter-AS traffic volume and the real-time property depend on the topology of the overlay network, which represents logical interconnections among peers. Therefore, to attain better performance in terms of the two measures, we should construct a proper topology of the overlay network. The topology is determined by *a peer selecting algorithm,* which is performed by a newly joining peer when it selects peers (providing peers) that provides the content data for it.

As conventional peer selection algorithms, Minimum-Hop (MH) [9] and Minimum-Depth (MD) [10-12] have already been proposed. MH aims at reducing inter-AS traffic volume. In MH, a newly joining peer selects a peer whose physical hop count from it is as small as possible as its providing peer. The physical hop count between two peers is defined as the number of ASes traversed on the content delivery path between them. On the other hand, MD aims at reducing content delivery delay. In MD, a newly joining peer selects a peer whose logical hop count is as small as possible as its providing peer. The logical hop count of a peer is defined as the number of peers or OSSes traversed on the content delivery path from an OSS to the peer. However, MH and MD do not take account of either inter-AS traffic volume or the real-time property of content. A CDN-based peer selection algorithm [13] tries to reduce inter-AS traffic volume by exploiting information available in content distribution networks (CDNs). In the algorithm, a newly joining peer estimates physical proximity to other peers with the help of CDN services, and selects a peer that has the closest proximity as its providing peer. However, this algorithm does not take account of the real-time property of content.

In order to reduce inter-AS traffic volume while keeping the real-time property of content, we have already proposed Minimum Physical Hop (MPH) [14, 15] as a peer selection algorithm. MPH aims at reducing both inter-AS traffic volume and content delivery delay by inheriting the peer selection policies from both MH and MD. The peer selection procedures of MPH are as follows. Firstly, to reduce inter-AS traffic volume, a newly joining peer selects a peer with the minimum physical hop count from it as its providing peer. Secondly, if there are several such candidates, to reduce content delivery delay, it selects a peer with the minimum logical hop count among the candidates as its providing peer. In addition, to strictly keep the real-time property of content, MPH sets an upper bound on logical hop count of candidates for providing peers, and a newly joining peer cannot select peers whose logical hop counts reach the upper bound as its providing peer.

In previous studies [14, 15], it was reported that, although MPH achieves low inter-AS traffic volume when the number of joining peers is small, it shows high inter-AS traffic volume when the number of joining peers is large. This reason is as follows. When we use MPH, candidates for providing peers with large logical hop counts tend to swarm in ASes without an OSS while those with small logical hop counts do in ASes with an OSS because of the peer selection policy of MPH. As the number of joining peers becomes large, the logical hop counts of the candidates for providing peers in every AS increase. As a result, in the former ASes, the logical hop counts of all the candidates for providing peers reach the upper bound frequently, and the newly joining peer cannot often find its providing peers in the AS, and consequently it has to select its providing peers in different ASes, which results in increased inter-AS traffic volume.

In this paper, to solve the problem of MPH, we propose Improved Minimum Physical Hop (IMPH) as a novel peer selection algorithm. One way to solve the problem is to distribute *top level peers,* which are peers with one logical hop count, to as many ASes as possible. In IMPH, when there are no candidates for providing peers in an AS, a newly joining peer in the AS gives high priority to selecting an OSS instead of the peer with the minimum physical hop count as its providing peer. As a result, it becomes a top level peer, and consequently we expect that the following peers in the AS successfully find its providing peer within the same ASes. In addition, because providing capacity of an OSS is limited, to prevent peers in ASes with an OSS from occupying the capacity, a newly joining peer in the ASes gives high priority to selecting a peer except an OSS within the same AS.

The rest of this paper is organized as follows. In Section 2, we describe the P2P live streaming system that we assume. In Section 3, we explain MPH and its problem. In Section 4, we propose IMPH. In Section 5, we show the evaluation results. In Section 6, we conclude our paper.

## 2. P2P LIVE STREAMING SYSTEM

Figure 1 depicts the P2P live streaming system assumed in this paper. The system consists of Origin Streaming Servers (OSSes), a peer control server, and peers. The OSSes produce original content data and provide the content data for peers. The peer control server maintains a list of peers that have already joined the system and provides the list for peers. The peers are the user terminals that have already joined the system, and relay the content data to other peers. Hereafter we call a peer that provides the content data for other peers *a providing peer* and a peer that receives the content data from other peers *a receiving peer*.

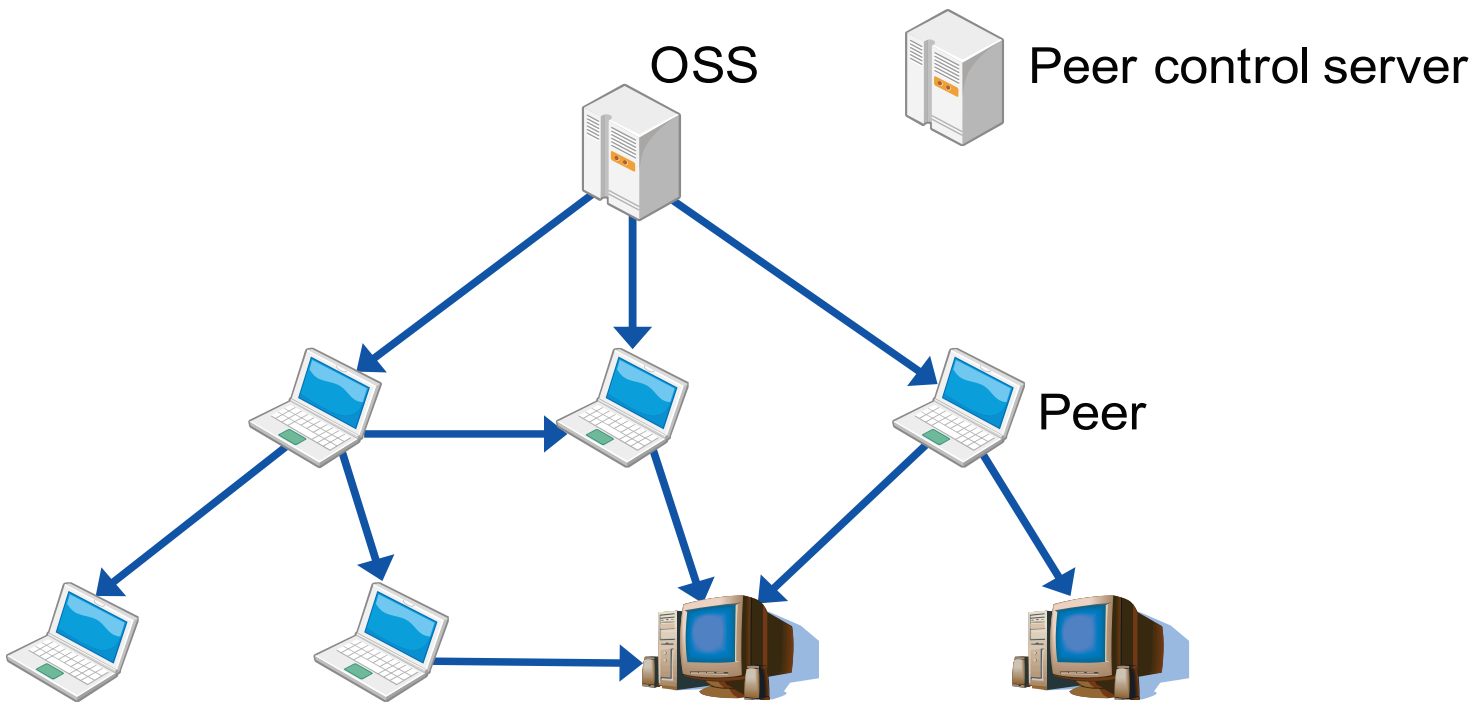

Figure 1. P2P live streaming system

A procedure for a peer to join the system is as follows. When a peer (newly joining peer) joins the system, it requires a list of candidates for providing peers of the peer control server (step 1). After getting the list, the newly joining peer selects providing peers from the list with *a peer selection algorithm* (step 2). Then, the newly joining peer sends out a delivery request to the selected providing peers (step 3). After receiving the content data from the providing peers, the newly joining peer can begin to view the content (step 4).

After joining the system, the newly joining peer needs to receive the streaming rate ($A$ Mbps) of the content data from the providing peers including OSSes in order to continue viewing the content. We assume that the streaming rate can be divided into $N$ minimum units and the rate of a minimum unit is $a$ Mbps. Accordingly, newly joining peers have to satisfy the following equation in order to continue viewing the content.

$$A = N \times a = (N_1 + N_2 + \cdots N_D) \times a \tag{1}$$

where $N_i$ is the number of minimum units provided by providing peer $i$ and $D$ is the number of providing peers.

While newly joining peers continue receiving the streaming rate of $A$ Mbps, they can become providing peers. The number ($M$) of minimum units that each providing peer can provide for other peers changes depending on its upload bandwidth. We call the number of minimum units *the providing capacity*. Providing peers have to satisfy the following equation.

$$M \times a = (M_1 + M_2 + \cdots M_U + S) \times a \tag{2}$$

where $M_j$ is the number of minimum units provided for receiving peer $j$, $U$ is the number of receiving peers, and $S$ is the remaining providing capacity.

Figure 2 depicts the network model of our P2P live streaming system. An overlay network represents interconnections among peers. An arrow in the overlay network shows delivery of the content data between the two peers. For example, in Fig. 2, peer 1 receives the content data from the OSS and provides it for peer 4. A physical network represents physical interconnections among ASes.

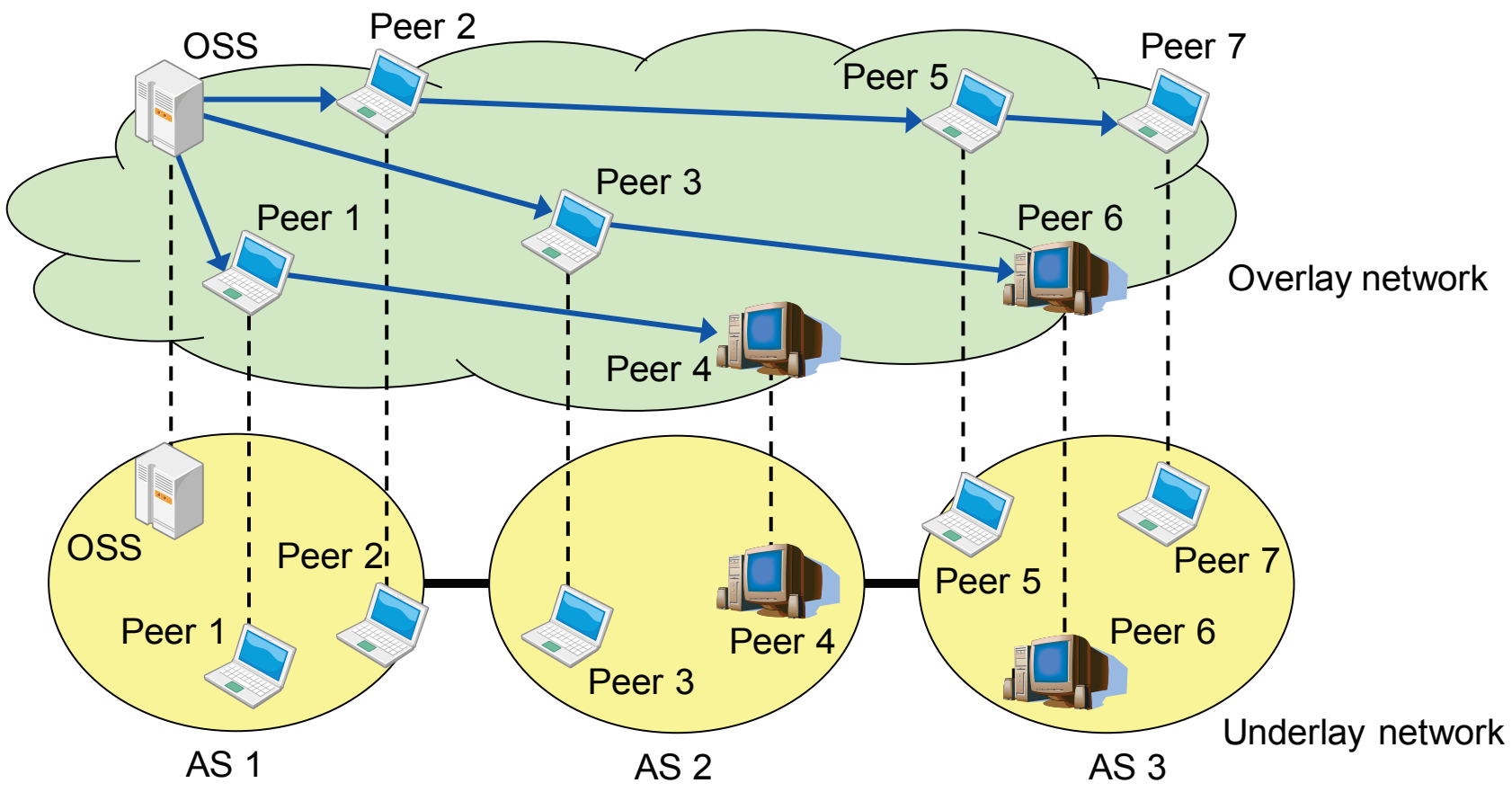

Figure 2. Overlay and underlay networks

## 3. Conventional Peer Selection Algorithm

We have already proposed Minimum Physical Hop (MPH) [14, 15] as a conventional peer selection algorithm. MPH aims at reducing inter-AS traffic volume and content delivery delay by inheriting peer selection policy from both Minimum-Hop (MH) [9] and Minimum-Depth (MD) [10-12].

In MPH, a newly joining peer determines its providing peers based on *physical hop counts* and *logical hop counts* of candidates for providing peers. A physical hop count between two peers is defined as the number of ASes traversed on the content delivery path between them. For example, in Fig. 2, the physical hop count between peer 1 and peer 4 is two. A logical hop count of a peer is defined as the number of peers or OSSes traversed on the content delivery path from an OSS to the peer. For example, in Fig. 2, the logical hop count of peer 1 is one and that of peer 4 is two.

In order to keep the real time property of the content, MPH introduces an upper bound ($H$) on logical hop counts of candidates for providing peers, namely, only the peers whose logical hop counts are smaller than $H$ can become candidates for providing peers. For example, in Fig. 2, if $H$ is set to three, peer 7 cannot become a candidate for a providing peer while the other peers and the OSS can become providing peers.

In MPH, in order to reduce inter-AS traffic volume, a newly joining peer selects a peer whose physical hop count from it is as small as possible as its providing peer. If there are multiple such candidates, in order to reduce content delivery delay, it selects the peer whose logical hop is as small as possible as its providing peer. This algorithm is described as follows in detail.

1. A newly joining peer ($P_{new}$) obtains the list of candidates for its providing peers (i.e., peers with remaining providing bit-rate greater than or equal to $a$ and logical hop counts smaller than $H$) from the peer control server.
2. $P_{new}$ sorts the list in ascending order of physical hop counts from it. If there are multiple candidates with the same physical hop counts from $P_{new}$, it sorts them in ascending order of logical hop counts.

3. $P_{new}$ selects the candidate one by one from the head of the list until the sum of the remaining providing bit-rate of the selected candidates reaches the streaming rate ($A$) of the content data. If enough candidates are found, $P_{new}$ successfully finishes the algorithm. Otherwise, $P_{new}$ fails to join the system and the algorithm is finished.

Figure 3 shows an example of peer selection by MPH. Suppose that 1) an OSS and all peers have non-zero providing bit-rate, 2) the upper bound ($H$) on the number of providing peers is two, and 3) peers 1 and 2 have already joined the system while peers 3, 4, 5, 6 and 7 newly join the system in this order.

We first focus on peer selection in AS 4, which does not have the OSS. When peer 3 newly joins the system in AS 4, peer 3 obtains the list of candidates: {OSS, 1, 2}. Then, it sorts the list into {2, OSS, 1} based on the physical hop counts from it and the logical hop counts of the candidates. Finally, peer 3 selects peer 2 as its providing peer. Similarly, peers 4 and 5 in AS 4 select peer 2 as their providing peers. Note that peers 4 and 5 do not select peer 3 as their providing peers because the hop count of peer 3 reaches two (=$H$) and peer 3 cannot become a providing peer to keep the real time property of content.

We next focus on peer selection in AS 1, which has the OSS. When peer 6 newly joins the system in AS 1, peer 6 obtains the list of candidates and sorts it into {OSS, 1, 2}. Finally, peer 6 selects the OSS as its providing peer. In the same way, peer 7 in AS 1 selects the OSS as its providing peer.

As shown in the above example, providing peers with small logical hop counts tend to swarm in ASes with an OSS while providing peers with large logical hop counts tend to swarm in ASes without an OSS in MPH. In the latter ASes, in order to keep the real time property of content, newly joining peers (peers 3, 4 and 5 in Fig. 3) often have to select providing peers (peer 2 in Fig. 3) in different ASes, which leads to increased inter-AS traffic volume.

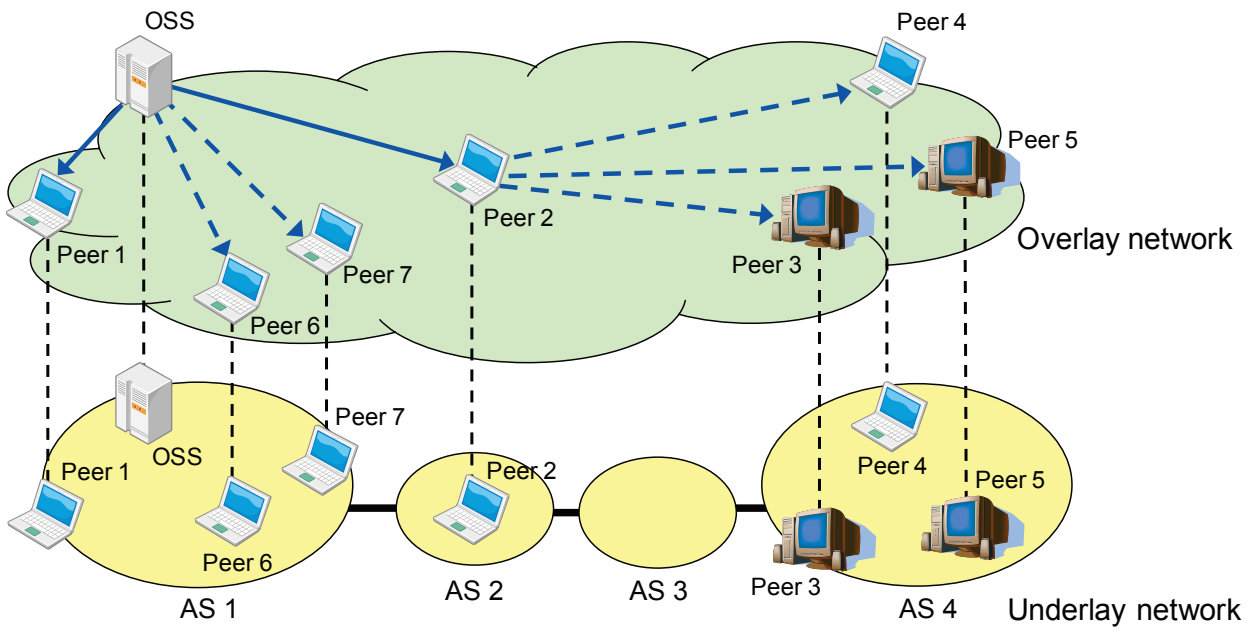

Figure 3. Peer selection by MPH

## 4. PROPOSED PEER SELECTION ALGORITHM

In order to solve the problem of MPH, we propose Improved Minimum Physical Hop (IMPH) as a novel peer selection algorithm. In IMPH, we try to decrease the inter-AS traffic volume by distributing *top level peers*, which are peers with one logical hop count, to as many ASes as possible. We expect that the distributed top level peer encourages the following peers within the same AS to select the peers within the same AS as their providing peers. This is because top level peers and their descendant peers tend to have small logical hop counts and newly joining peers within the same AS tend to select such the peers as their providing peers in the peer selection policy of MPH. In addition, in order to encourage top level peers to distribute to many ASes, we prevent the top level peers from swarming in ASes with an OSS by forcing a newly joining peer in ASes with an OSS to select a peer except the OSS as its providing peer.

In IMPH, firstly, in order to prevent peers in ASes with an OSS from occupying the providing capacity of OSS, a newly joining peer selects the peer with the minimum logical hop count except the OSS within the same AS. Secondly, if the AS does not have enough providing peers to satisfy the newly joining peer's request, in order to distribute top level peers to as many ASes as possible, the newly joining peer selects the OSS with the minimum physical hop count from it as its providing peer. Thirdly, if the newly joining peer still cannot obtain the enough content data, the newly joining peer selects its providing peers in the same way as MPH.

The algorithm of IMPH is obtained by replacing step (2) of MPH in Section 3 with the following steps.

(2-1) $P_{new}$ divides the list into the following three lists; 1) $L_1$: the list of candidates for providing peers except OSSes in the AS having $P_{new}$, 2) $L_2$: the list of all OSSes, and 3) $L_3$: the list of all the remaining candidates for providing peers (i.e., the peers that are not in the AS having $P_{new}$ and are not OSSes).
(2-2) $P_{new}$ sorts $L_1$ in ascending order of logical hop counts.
(2-3) $P_{new}$ sorts $L_2$ in ascending order of physical hop counts from it.
(2-4) $P_{new}$ sorts $L_3$ in ascending order of physical hop counts from it. If there are multiple candidates with the same physical hop counts from it, $P_{new}$ sorts them in ascending order of logical hop counts.
(2-5) $P_{new}$ concatenates $L_1$, $L_2$ and $L_3$ into the single list in this order.

The time complexity of IMPH is analyzed as follows. We denote the number of candidates for providing peers in the list by $n$. The time complexity of step (2-1) is $O(n)$ because we classify all the candidates into one of the three lists one by one. The time complexity of each of steps (2-2) to (2-4) is $O(n\log n)$ because we sort at most $n$ candidates in each step. Therefore, the run time complexity for IMPH is $O(n\log n)$.

We explain an example of peer selections by IMPH in Fig. 4. The preconditions in Fig. 4 are the same as those in Fig. 3.

We first focus on peer selection in AS 4, which does not have the OSS. Peer 3, which is a newly joining peer, obtains the list of candidates {OSS, 1, 2}. Then peer 3 divides the list into three lists: $L_1$ = {null}, $L_2$ = {OSS}, and $L_3$ = {1, 2}. Peer 3 next sorts each of the lists as follows: $L_1$ = {null}, $L_2$ = {OSS}, $L_3$ = {2, 1}. Then, peer 3 concatenates the three lists into the single list: {OSS, 2, 1}. Finally, peer 3 selects the OSS as its providing peer. Note that peer 3 becomes a top level peer, consequently a top level peer is successfully distributed to AS 4. Next, peer 4 obtains the list of candidates {OSS, 1, 2, 3}. Then, peer 4 divides the list into three lists: $L_1$ = {3}, $L_2$ = {OSS}, and $L_3$ = {1, 2}. Peer 4 sorts each of the lists and concatenates them into the single list: {3, OSS, 2, 1}. Finally, peer 4 selects peer 3 as its providing peer. Similarly, peer 5 selects peer 3 as its providing peer. Note that peers 4 and 5 succeed in finding their providing peers within the same AS and the inter-AS traffic volume is decreased thanks to distributing a top level peer (peer 3 in Fig. 4) to AS 4.

We next focus on peer selection in AS 1, which has the OSS. Peer 6 obtains the list of candidates and divides the list into three lists: $L_1$ = {1}, $L_2$ = {OSS}, and $L_3$ = {2, 3, 4, 5}. Peer 6 sorts each of the lists and concatenates them into the single list: {1, OSS, 2, 3, 4, 5}. Finally, peer 6 selects peer 1 as its providing peer. Similarly, peer 7 selects peer 1 as its providing peer. Note that, in AS 1, only peer 1 selects the OSS as its providing peer and the providing capacity of the OSS is saved for distributing top level peers to other ASes.

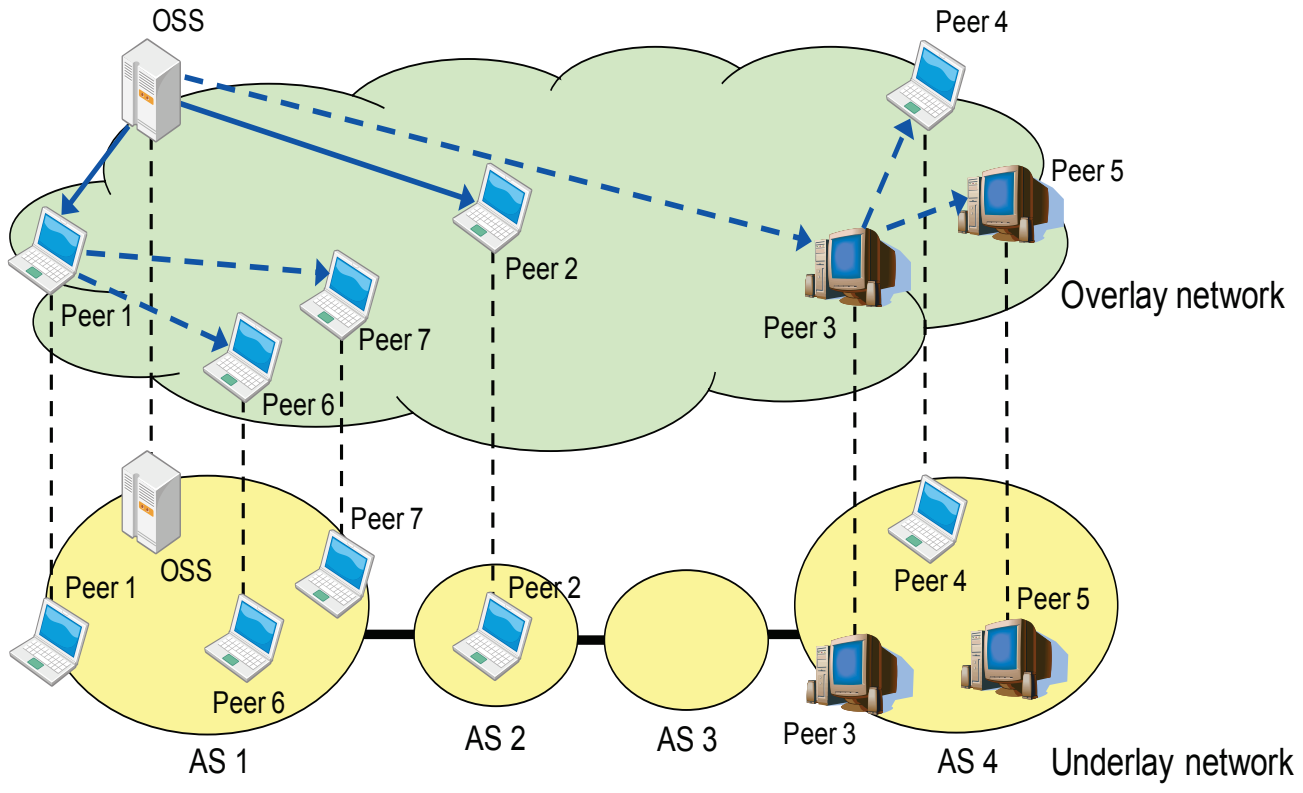

Figure 4. Peer selection by IMPH

# 5. PERFORMANCE EVALUATION

## 5.1. SIMULATION MODEL

We evaluate the performances of IMPH and MPH with simulations. Tables 1 shows the parameter settings in the simulations. Note that only the providing rate of peer $i$ changes on a per-peer basis in one simulation run while other parameter values are fixed throughout one simulation run. Table 2 shows the AS topology parameters. The topology is made with BRITE [16] and based on a BA model [17]. The model has a scale-free character, where most of ASes have a few degrees and a few ASes have many degrees [18].

We evaluate the performance of algorithms by using *congestion degree*, which means the relative volume of inter-AS traffic. Congestion degree is defined by the following equation.

$$C = \frac{\sum_{0 \le i \le k} \sum_{0 \le j \le k} \lambda_{ij} h_{ij}}{\sum_{0 \le i \le k} \sum_{0 \le j \le k} \lambda_{ij}} \quad (3)$$

where $i$ and $j$ are indexes of peers or OSSes, $\lambda_{ij}$ is the bit-rate of content data from $i$ to $j$, $k$ is the number of joining peers and OSSes and $h_{ij}$ is the physical hop count from $i$ to $j$. The denominator is the total traffic volume and the numerator is the total traffic volume weighted by physical hop count. Thus, the minimum ideal congestion degree is one.

Table 1. Simulation parameters.

| Parameter | Value |
|---|---|
| Streaming rate ($A$) | 1 [Mbps] |
| Minimum unit ($a$) of streaming rate | 256 [kbps] |
| Number of OSSs | 10 |
| Providing rate of OSS | 100 [Mbps] |
| Deployment of OSSs | Random |
| Upper bound ($H$) on logical hop counts of providing peers | 3, 4, 5 |
| Providing rate of peer $i$ | Uniform distribution on the interval [$a$, $M_{max}$ * $a$] |
| Upper bound ($M_{max}$) on the number of the minimum units that each peer can provide | 20, 40 |

Table 2. AS topology parameters.

| Parameter | Value |
|---|---|
| Number of ASes | 500 |
| Diameter of the topology | 7 [hops] |
| Average distances to other ASes | 3.73 [hops] |
| Maximum degree | 59 |
| Minimum degree | 2 |
| Average degree | 3.91 |

## 5.2. RESULTS

Figures 5 through 10 depict congestion degrees of IMPH and MPH when the *n*th peer joins the system (value *n* of x-axis corresponds to the generation of the *n*th peer).

Congestion degrees of both IMPH and MPH sharply decrease when the number of joining peers is smaller than around 2,000. This is because candidates for providing peers spreads throughout all the ASes as the number of joining peers increases, consequently the following peers in all the ASes can find their providing peers within the same ASes.

The congestion degrees of MPH gradually increase when the number of joining peers becomes larger than a certain value. For example, the congestion degree of MPH begins to increase when the number of joining peers reaches around 2,000 and 3,000 in Figs. 5 and 7, respectively. This is because newly joining peers often fail to find their providing peers within the same AS and have to select the peers in other ASes as their providing peers in ASes without an OSS.

On the other hand, the congestion degrees of IMPH continue decreasing until when more peers join the system compared to MPH due to succeeding in distributing top level peers to many ASes. For example, the congestion degree of IMPH keeps decreasing until when the number of joining peers is around 5,500 and 15,000 in Figs, 5 and 7, respectively.

The congestion degree of IMPH can be higher than that of MPH when the number of joining peers is smaller than a certain value (e.g., 1,300 in Fig. 5). This is because, at the initial stage of the system, IMPH causes the inter-AS traffic volume in order to distribute top level peers to all the ASes while MPH avoid such inter-AS traffic volume by forcing a newly joining peer to simply select the peer whose physical hop count from it is the minimum as its providing peer.

Except for the aforementioned situation (i.e., the number of joining peers is higher than a certain value), the congestion degree of IMPH is lower than that of MPH. The congestion degree of IMPH is at a maximum of 50%, 63%, 56%, 64%, 50% and 63% lower than that of MPH in Figs. 5, 6, 7, 8, 9 and 10, respectively.

It is reported that the number of joining peers is tens of thousands of peers in an actual system [2]. In addition, because inter-AS traffic volume becomes larger as the number of joining peers increases, it is important to suppress the volume when the number is large. Therefore, we conclude that IMPH is more suitable than MPH in an actual P2P live streaming environment.

## 6. CONCLUSION

In this paper, we have proposed IMPH as a novel peer selection algorithm for decreasing inter-AS traffic volume in P2P live streaming. IMPH tries to reduce the inter-AS traffic volume by distributing top level peers to as many ASes as possible. We have evaluated the performance of IMPH with that of MPH by simulation. Numerical examples have shown that 1) IMPH achieves at a maximum of 64% lower inter-AS traffic volume than MPH when the number of joining peers is large and 2) IMPH shows higher inter-AS traffic volume than MPH when the number is small.

One of our future work is to propose an overlay reconstruction algorithm in order to cope with the situation where peers frequently join and leave the system.

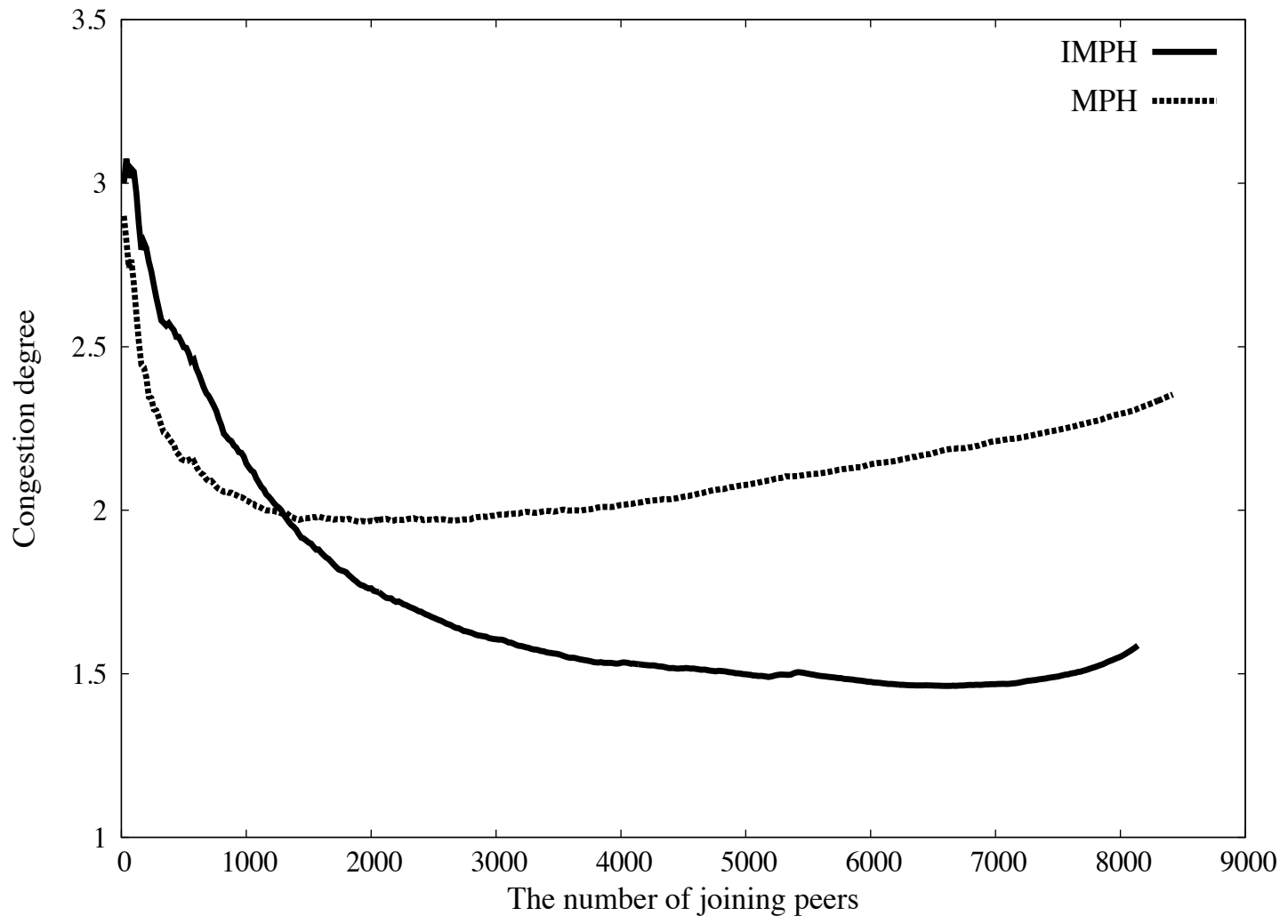

Figure 5. Congestion degree ($H$=3, $M_{max}$ = 20)

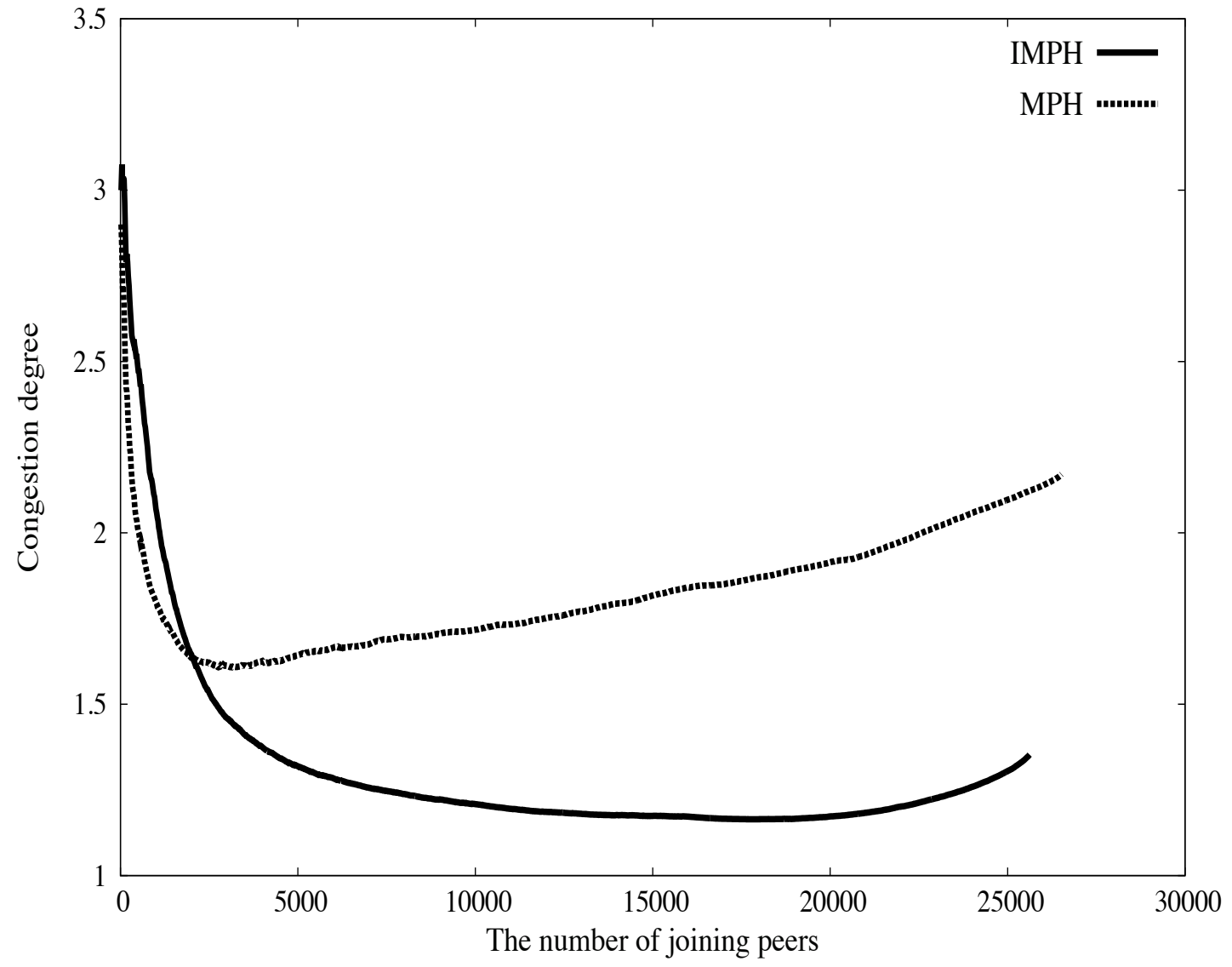

Figure 6. Congestion degree ($H$= 3, $M_{max}$ = 40)

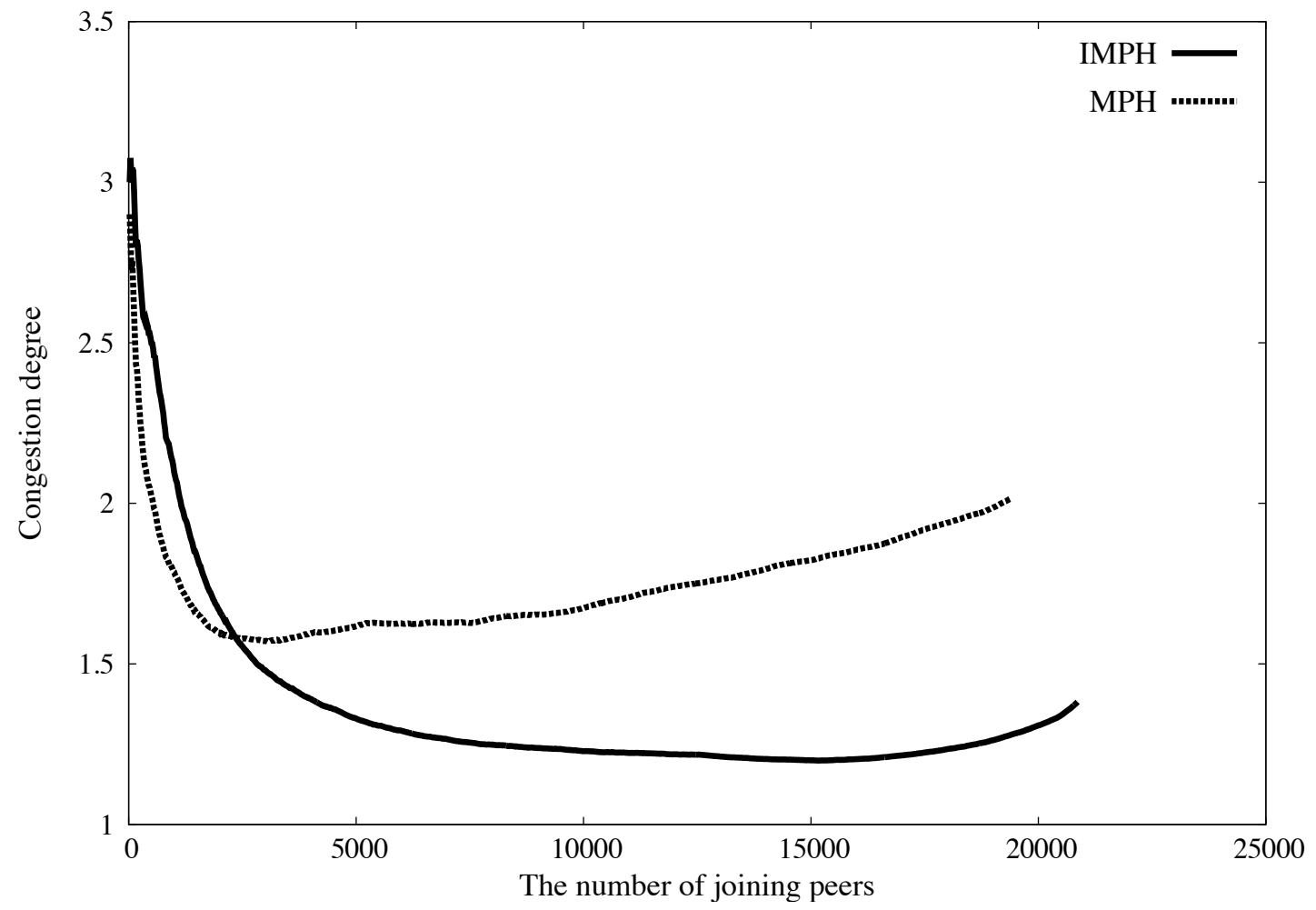

Figure 7. Congestion degree ($H$= 4, $M_{max}$ = 20)

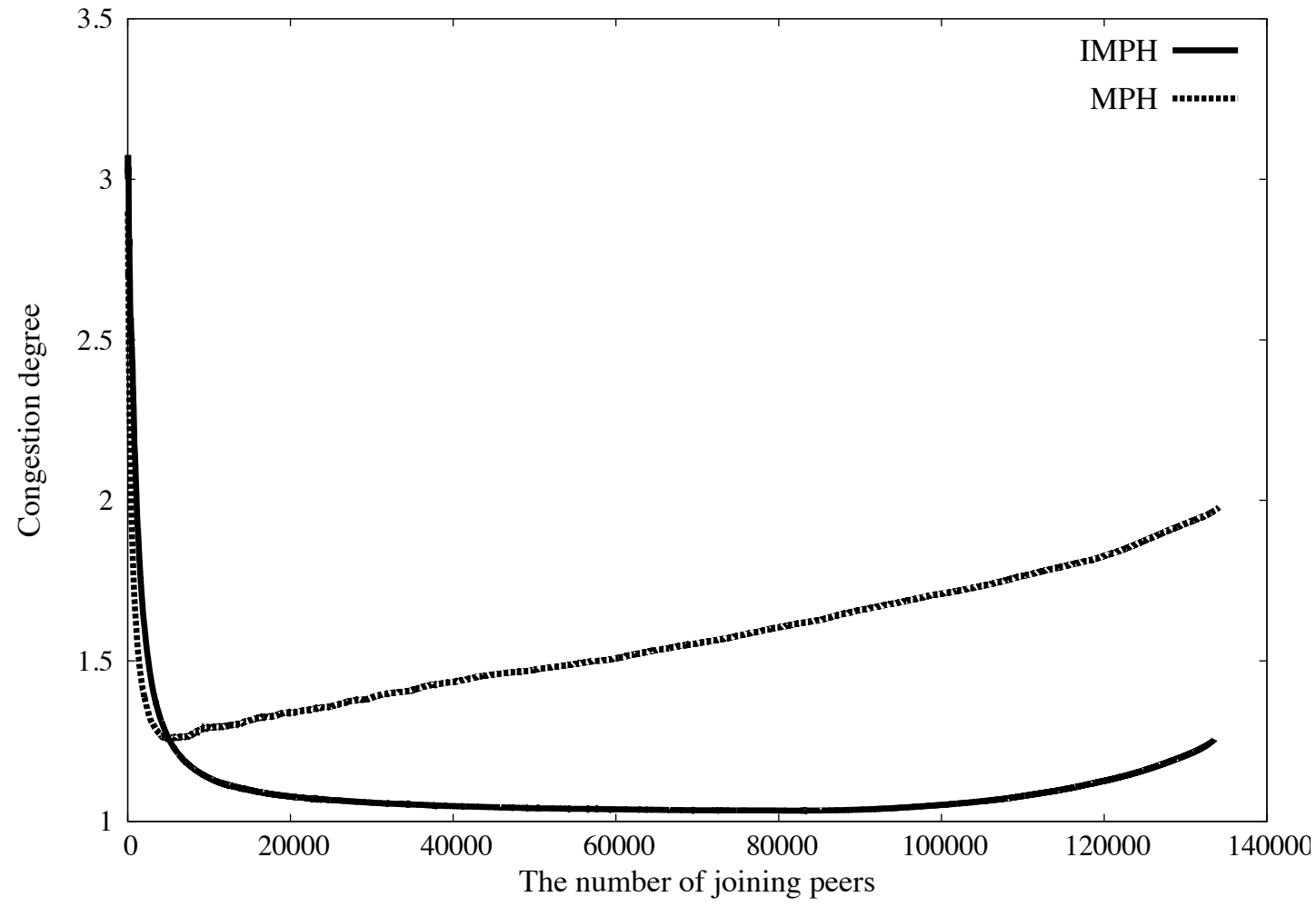

Figure 8. Congestion degree ($H$= 4, $M_{max}$ = 40)

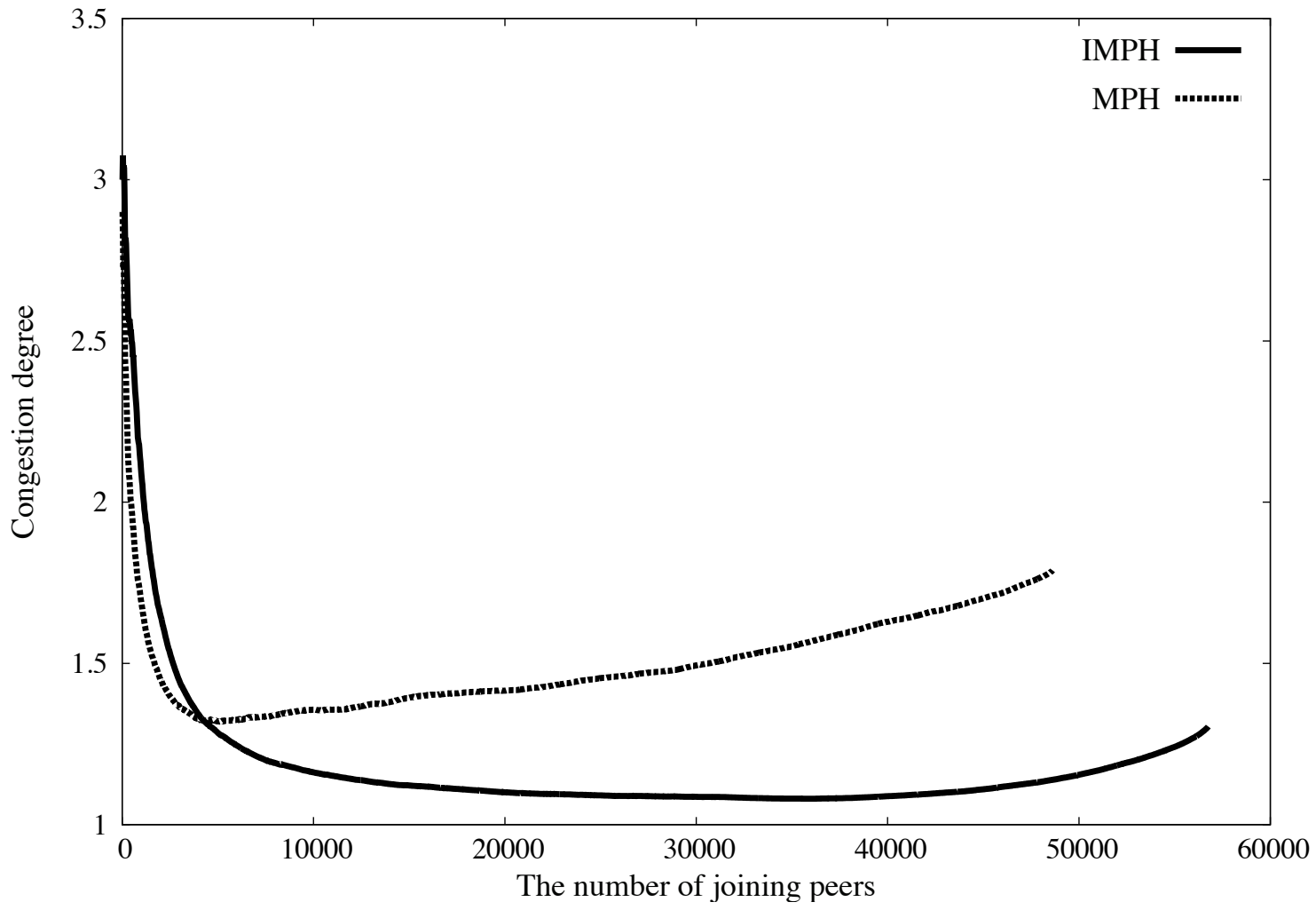

Figure 9. Congestion degree ($H$= 5, $M_{max}$ = 20)

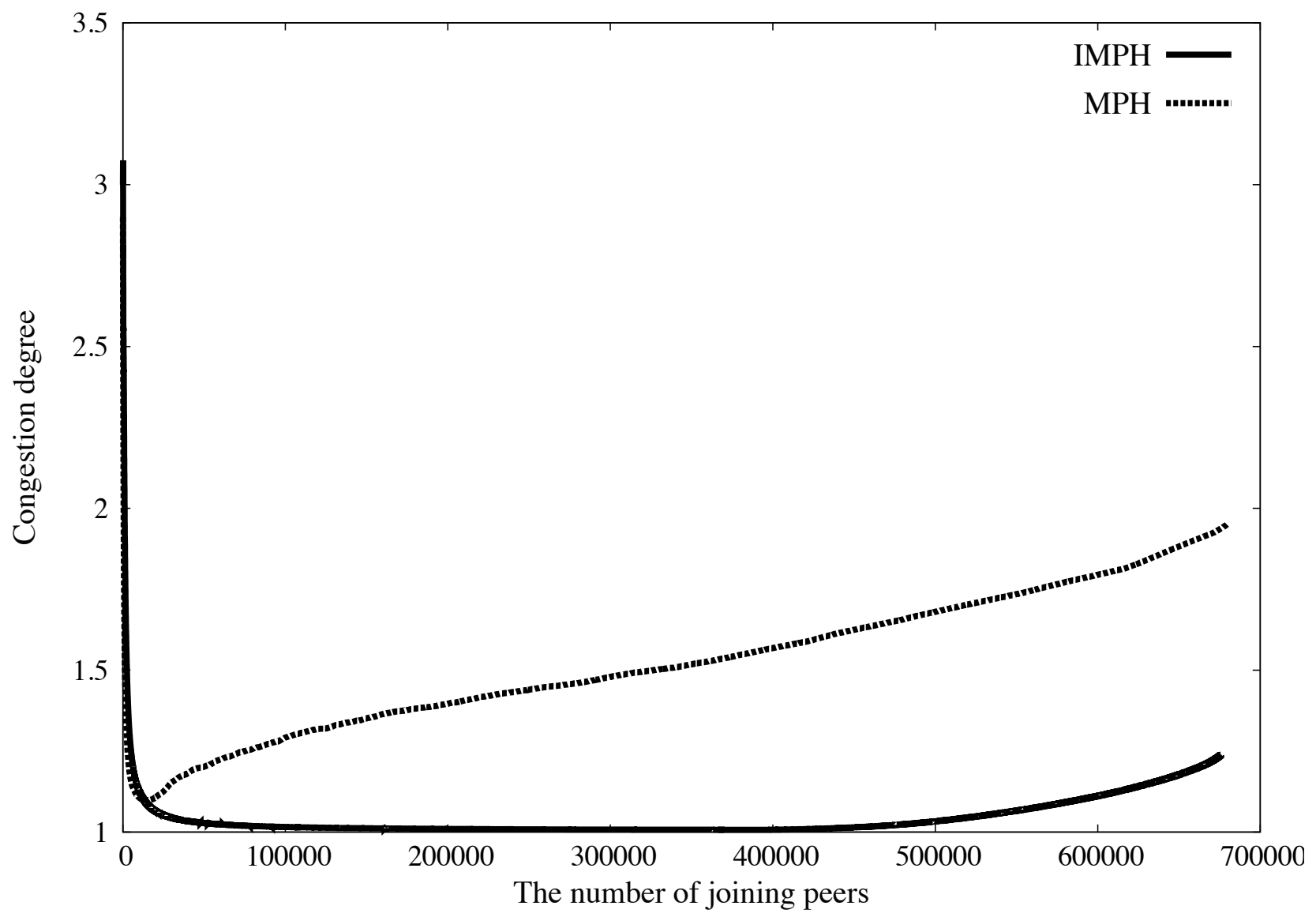

Figure 10. Congestion degree ($H$= 5, $M_{max}$ = 40)

## AUTHORS

**Yukinobu Fukushima** received his B.E., M.E. and Ph.D. degrees from Osaka University, Japan, in 2001, 2003 and 2006. He is currently an assistant professor of the Graduate School of Natural Science and Technology, Okayama University. His research interest includes optical networking, P2P live streaming and cloud computing. He is a member of IEEE, ACM, OSA, and IEICE.

**Kazuki Kobayashi** received his B.E. and M.E. degrees from Okayama University, Japan, in 2010 and 2012.

**Tokumi Yokohira** received the B.E., M.E. and Ph.D. degrees in information and computer sciences from Osaka University, Osaka, Japan, in 1984, 1986 and 1989, respectively. From April 1989 to May 1990, he was an assistant professor in the Department of Information Technology, Faculty of Engineering, Okayama University, Okayama, Japan. From May 1990 to December 1994 and from December 1994 to March 2000, he was a lecturer and an associate professor, respectively, in the same department. From April 2000 to June 2003, he was an associate professor in the Department of Communication Network Engineering of the same faculty. Since July 2003, he has been a professor in the same department. His present research interests include performance evaluation and improvement of computer networks and communication protocols, design algorithm of optical networks and network securities. He is a member of the IEEE Communication Society, the Institute of Electronics, Information and Communication Engineers and Information Processing Society of Japan.